\RequirePackage{fix-cm}
\documentclass[twocolumn]{svjour3}          
\smartqed  
\usepackage{graphicx}
\usepackage{balance}
\usepackage{microtype}
\usepackage{subfigure}
\usepackage{tabu}                      
\usepackage{booktabs}                  
\usepackage{hyperref}
\usepackage{amssymb}
\journalname{Virtual Reality}
\usepackage{xcolor}
\newcommand{\changed}[1]{#1}

\begin{document}


\title{Controlling Camera Movement in VR Colonography}


\author{Soraia F. Paulo \and Daniel Medeiros \and Daniel Lopes \and Joaquim Jorge
}


\institute{Soraia F. Paulo, Daniel S. Lopes,  Joaquim Jorge \at
              INESC-ID Lisboa, Instituto Superior Técnico, ULisboa \\
              Av. Prof. Dr. Cavaco Silva, 2744-016, Porto Salvo, Portugal\\
              \email{jaj@inesc-id.pt, jorgej@tecnico.ulisboa.pt}
           \and
           Daniel Medeiros \at
              University of Glasgow, United Kingdom
}

\date{Received: 18 June 2021 / Accepted: 21 December 2021}

\maketitle

\begin{abstract}
Immersive Colonography allows medical professionals to navigate inside the intricate tubular geometries of subject-specific 3D colon images using Virtual Reality displays. Typically, camera travel is performed via Fly-Through or Fly-Over techniques that enable semi-automatic traveling through a constrained, well-defined path at user-controlled speeds. However, Fly-Through is known to limit the visibility of lesions located behind or inside \changed{haustral folds}. 
At the same time, Fly-Over requires splitting the entire colon visualization into two specific halves. In this paper, we study the effect of immersive Fly-Through and Fly-Over techniques on lesion detection and introduce a camera travel technique that maintains a fixed camera orientation throughout the entire medial axis path. While these techniques have been studied in non-VR desktop environments, their performance is not well understood in VR setups. We performed a comparative study to ascertain which camera travel technique is more appropriate for constrained path navigation in Immersive Colonography \changed{and validated our conclusions with two radiologists.} To this end, we asked 18 participants to navigate inside a 3D colon to find specific marks. Our results suggest that the Fly-Over technique may lead to enhanced lesion detection at the cost of higher task completion times. 
Nevertheless, the Fly-Through method may offer a more balanced trade-off between speed and effectiveness, whereas the fixed camera orientation technique provided seemingly inferior performance results. Our study further provides design guidelines and informs future work.


\end{abstract}

\section{Introduction}
\label{intro}

Colorectal cancer (CRC) is the second leading cause of cancer-related death in the western world, with an estimated 1.4 million new cases every year worldwide, half of which end in death~\cite{Ferlay2015}. Computed Tomography Colonography (CTC) is an imaging technique that has been widely adopted for colonic examination for diagnostic purposes. Still, the colon is an organ with several inflections and numerous colonic haustral folds along its extension, making navigation inside CTC 3D models a hard task~\cite{yao2010reversible}. 

While analyzing CTC content, radiologists work in the standard workstation, i.e., desktop, monitor, mouse, and keyboard. However, using a 2D display to analyze 3D structures can lead to missing information~\cite{mirhosseini2014}. As conventional systems rely on 2D input devices and stationary flat displays, clinicians often struggle to obtain the desired camera position and orientation, which requires several cumulative rotations, making it hard to perceive the colon structure in 3D. 
To visualize such anatomically complex data, the immersion and freedom of movement afforded by VR systems bear the promise to assist clinicians in improving 3D reading, namely enabling more expedite diagnoses~\cite{Schuchardt07}. 

Travel is considered the most basic and essential component of the VR experience, which is responsible for changing the user's viewpoint position and rotation in a given direction~\cite{bowman1997travel}.
Due to the complexity of large virtual environments, several authors apply travel techniques that rely on path planning, i.e., path-based or path-constrained travel. In this family of techniques, the user follows a previously defined path, where users can still control speed, viewpoint direction, and local deviation, so that they can locally explore the virtual environment ~\cite{elmqvist2006navigation,elmqvist2007tour}. 
Path-based travel can also be done automatically, where both the path and the movement are predefined to create a smooth navigation experience. Such features are welcome for virtual endoscopy applications~\cite{bartz2005virtual,he2001reliable,huang2005teniae}.

Given the complexity of the colon's structure, travel follows a semi-automatic procedure that relies on centerline estimation to constrain the direction of movement. Still, users can control speed. The most conventional way of CTC travel consists of the Fly-Through technique ~\cite{hong19953d}, where camera orientation follows the centerline's direction. Nonetheless, VR could enable more natural means of travel by decoupling camera orientation from the direction of movement, in the sense that relative orientation can differ from the centerline's direction. That is the case of the Fly-Over technique, where relative orientation is perpendicular to the centerline's direction~\cite{hassouna2006flyover}. Although these techniques are commonly used in conventional setups, they have yet to be thoroughly investigated in VR settings. Our work focuses on camera travel as a critical component of surveying and identifying pathological features in CTC datasets. 
The semi-automatic nature of the process, combined with the abrupt direction changes caused by the complexity of the colon's structure, may cause unwanted side-effects due to the difference between camera orientation in the virtual world and the user's real orientation~\cite{Robinett1992}.
We propose an Elevator technique to overcome this issue, where camera orientation changes to match the user's actual orientation. Using an immersive colonoscopy prototype  ~\cite{lopes2018interaction} we studied camera control techniques and their effectiveness on comprehensive landmark identification in order to address the following question:
\emph{Which of the considered travel techniques is the most suitable to navigate inside the 3D reconstructed model of the colon?}

\section{Related Work}
\label{relatedWork}

Navigation inside colon structures is a non-trivial and challenging task task to perform. The Fly-Through technique has been widely adopted since it was first proposed by Lichan Hong~\cite{hong19953d}. Radiologists prefer this type of visualization due to its similarities with conventional colonoscopy, which includes dealing with the same limitations. While moving in a given direction, lesion visibility is limited to the colorectal tissue exposed to the normal of the viewing camera, which may lead to missing significant lesions. In order to address this and reduce redundancy, colorectal flattening proposed mapping the colon's cylindrical surface to a rectangular plane to create a complete virtual view of the colon~\cite{Haker2000NondistortingFM}. Nonetheless, flattening algorithms are prone to error and require additional training to understand such 2D representation of the colon~\cite{wang2015novel}. The unfolded cube projection proposed projecting all views of the colon on the inside of a cube using six camera normals that move together along the centerline~\cite{vos2003three}. Even though unfolding the cube enhances lesion visibility, scanning all sides is a time-consuming task. 
Fly-Over is another visualization technique that tries to solve Fly-Through's limitations. In this case, the colon is divided into two unique halves by the centerline, each with a virtual camera~\cite{hassouna2006flyover}. This method enables perpendicular perspective, producing increased surface coverage (99\% of surface visibility in one direction vs. 93\% in Fly-Through in two navigation directions) and equally good sensitivity. Despite the Fly-Over's positive impact, current CTC software, such as the V3D-Colon\footnote{\textit{http://www.viatronix.com/ct-colonography.asp}}, syngo.CT Colonography\footnote{Siemens Healthineers, 2017.~\textit{syngo.CT Colonography}} only include Fly-Through, flattening, and the unfolded cube visualization techniques since the use of the Fly-Over is restricted for patent reasons. Still, these techniques suffer from using a conventional 2D interface to interact with a 3D model. 

Since its inception, Virtual Reality (VR) has found applications across the medical domain, namely in medical education~\cite{codd2011virtual}, surgical planning and training tasks \cite{de2011progress,vosburgh2013surgery,shanmugan2014virtual}. More recently, VR has also been applied to diagnosis~\cite{king2016immersive,sousa2017vrrrroom,wirth2018evaluation}, where being able to sift through large and complex image datasets is crucial to producing insightful and complete results. Thus, controlling viewing position and orientation in expedited yet precise manners could potentially affect significant medical decisions.

\changed{
Differently from commonly used locomotion techniques that often employ floor-constrained or 6DoF travel metaphors~\cite{medeiros2019magic}, locomotion in tubular anatomical structures requires constrained navigation through a pre-computed path~\cite{chaudhuri2004efficient,cheng2014augmented,noser2003automatic}.
Navigating inside the human body has been previously studied in a variety of different procedures, such as bronchoscopy~\cite{aguilar2017rrt} and angioscopy~\cite{haigron2004depth}.
Such procedures rely on constraining the locomotion mainly along the center-axis of the generated 3D model, with varying forms of camera orientation.
These structures are very complex and often need the path to be computed interactively~\cite{aguilar2017rrt,haigron2004depth}, which makes the path-planning procedure more critical than the travel technique used.
Colon structures, on the other hand, are not as ramified but contain haustral folds and inflections that abruptly change the direction of movement while traveling~\cite{huang2006synchronous}, which can promote unwanted side-effects, such as cybersickness and disorientation. As a result, this could highly hinder physicians chances of making the correct diagnosis in an efficient and effective way~\cite{pareek2018survey,venson2016medical}.
}

Considering the advantages of VR to diagnostic imaging, especially improved camera control, freedom of movement, 3D perception, and enhanced scale, two groups started to explore the immersive CTC experience. First, Mirhosseini et al. investigated a CAVE (Cave Automatic Virtual Environment) in which the gastrointestinal walls were projected onto the room walls~\cite{mirhosseini2014}. Despite suggesting potential improvement in reducing examination time and enhancing accuracy, this type of setup would be unrealistic in a natural clinical setting. More recently, Mirhosseini et al. proposed an immersive CTC system that leverages VR's advantages to improve lesion detection~\cite{mirhosseini2019immersive} while still relying on 2D interaction techniques. Similarly, Randall et al. explored an immersive VR prototype, obtaining encouraging feedback regarding overall faster diagnosis~\cite{randall2015oculus}. However, none of these works focus on camera travel, or explore a technique other than Fly-Through.

\section{Immersive Navigation of a 3D Virtual Colon}
\label{approach}

We developed an interactive VR system to assist 3D immersive navigation of subject-specific colon models enabling travel via Fly-Through, Fly-Over or Elevator camera modes~\cite{lopes2018interaction}.

\subsection{3D Data}

\begin{figure}[b]
    \centering
    \includegraphics[width=\columnwidth]{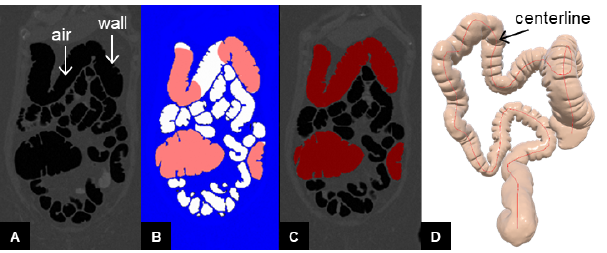}
    \caption{3D colon reconstruction: (A) original CTC image; (B) global threshold image with two active contours (red); (C) segmented colon overlapped with CTC image; (D) reconstructed 3D model with centerline}
    \label{fig:colonSegmentation}
\end{figure}

We used a single CTC dataset from \textit{The Cancer Imaging Archive} \cite{tciaCTC} (subject ID \textit{CTC-3105759107}) and reconstructed the 3D model using a freeware image-based geometric modeling pipeline~\cite{ribeiro20093} (\autoref{fig:colonSegmentation}).

The high contrast between luminal space (air: black) and colon luminal surface (colon wall: light grey) facilitates 3D reconstruction (\autoref{fig:colonSegmentation}(A)). Firstly, the 3D colon structure is segmented using the active contours method based on region competition (\autoref{fig:colonSegmentation}(B-C)), which depends on the intensity values estimated via a global threshold filter (ITK-SNAP 3.6). Secondly, a 3-D surface mesh of the segmented data is generated using marching cubes. Thirdly, undesired mesh artifacts were attenuated through a cycle of smoothing and decimating operations (ParaView 5.3.0) and exported into a *.ply (ASCII) file. Finally, the mesh file was converted to *.obj (Blender 2.78) and imported into Unity. To compute the 3D centerline of the colon mesh, we used the algorithm proposed by Tagliasacchi et al.~\cite{tagliasacchi2012mean} which solves the 3D mesh skeletonization problem by resorting on mean curvature flow (\autoref{fig:colonSegmentation}(D)).%

\subsection{\changed{Interaction Design}}

\begin{figure}[b]
\centering
\includegraphics[width=.65\columnwidth,trim=0 0 0 10, clip]{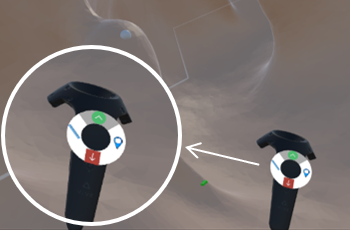}
\caption{Touchpad controller indicating the direction of movement: pressing the green arrow button the user moves forward, while pressing the red arrow the user moves backwards.\changed{To tag a lesion, the user must press the marker button, confirming with the controller's trigger button }}
\label{fig:interaction}
\end{figure}

\begin{figure}[t]
\centering
\includegraphics[width=.65\columnwidth]{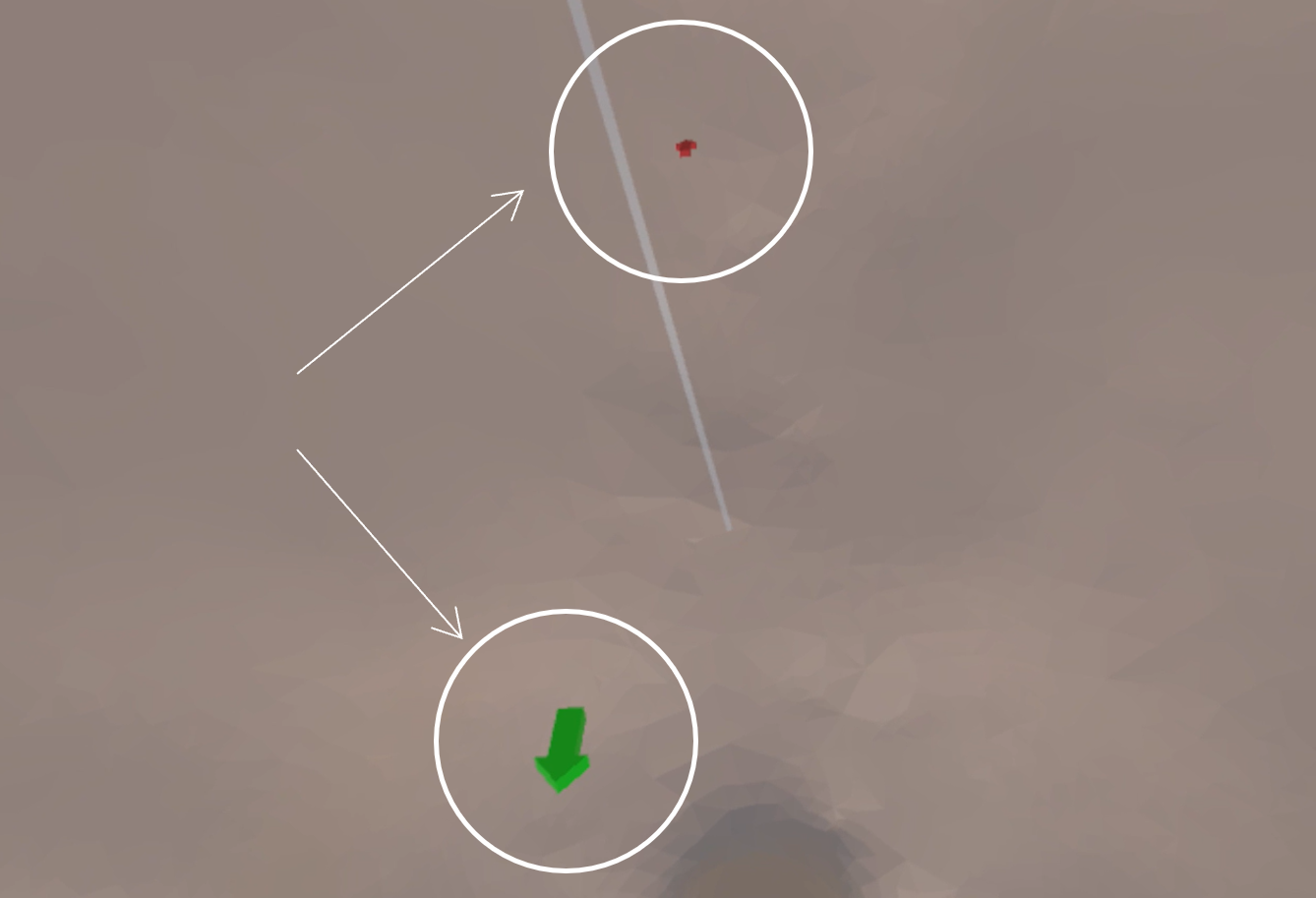}
\caption{\changed{Arrows indicating the direction of movement: antegrade (green) and retrograde (red).}}
\label{fig:arrows}
\end{figure}

\changed{Following conventional CTC practices, users are placed inside the colon and navigate from the rectum towards the cecum and \textit{vice-versa}.} All travel techniques follow a predefined centerline, i.e., follow the same path and use the same input (touchpad) to indicate the direction of movement (forward or backwards) at a constant speed (\autoref{fig:interaction}). \changed{By default, the user is anchored to the centerline to avoid unwanted intersections against the colon walls, while they can freely move their heads and/or body to look around and behind the virtual colon's tubular structure. However, users can opt to step away from the centerline by physically walking towards the colon wall and reach the lumen limits to better examine local features. After exploring the colon wall, users can reposition themselves by moving back towards the centerline. To assist navigation, two arrows pointing in opposite directions are placed in front of (green: antegrade) and behind (red: retrograde) the user accompanying the centerline (\autoref{fig:arrows}).}%

\changed{Navigational and diagnostic tools are managed through HTC Vive controllers (\autoref{fig:interaction}). A menu appears every time the touchpad is activated and tools are displayed by pressing the corresponding widget buttons. In this work, only the dominant hand controller was used to handle both the direction of movement and tagging tasks.}

\subsection{Immersive Camera Travel Techniques}

We considered three camera travel techniques which allowed users to inspect the colon model and navigate inside the luminal space: Fly-Through, Fly-Over and Elevator. 

\begin{figure}[t]
\centering
\subfigure[Fly-Through]{
\includegraphics[width=\columnwidth]{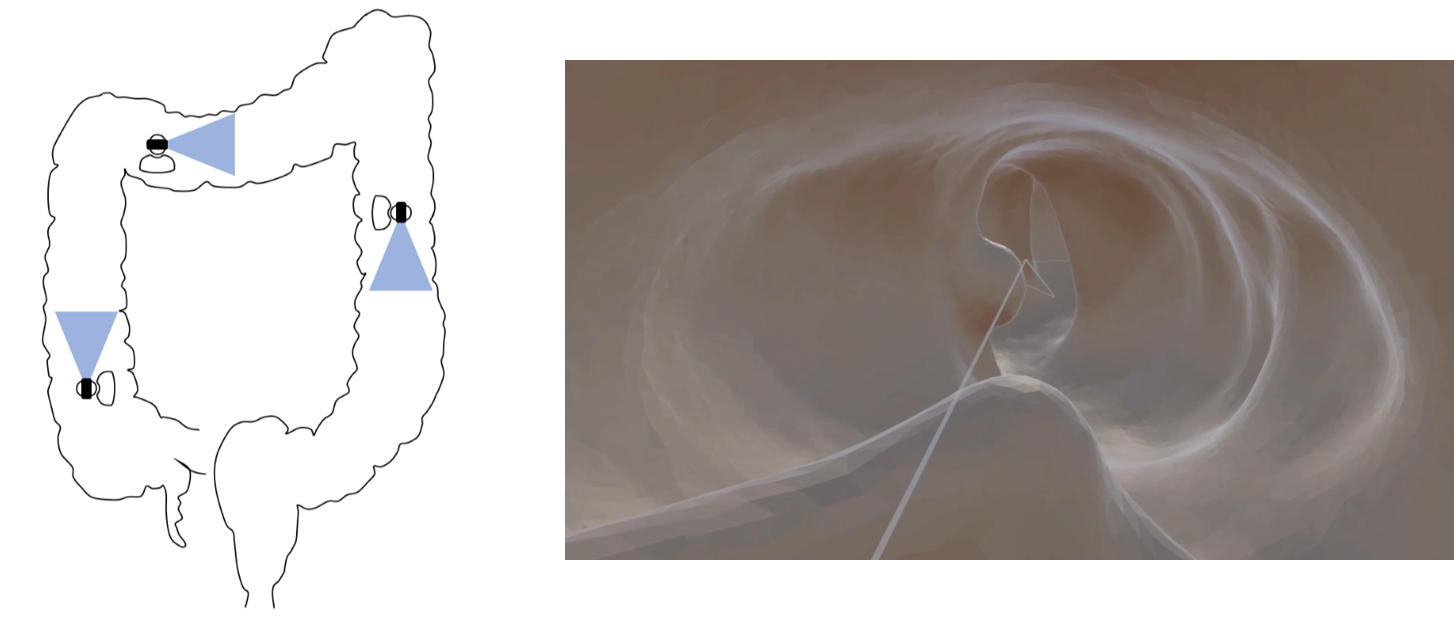}%
\label{fig:flyThrough}
}
\subfigure[Fly-Over]{
\includegraphics[width=\columnwidth]{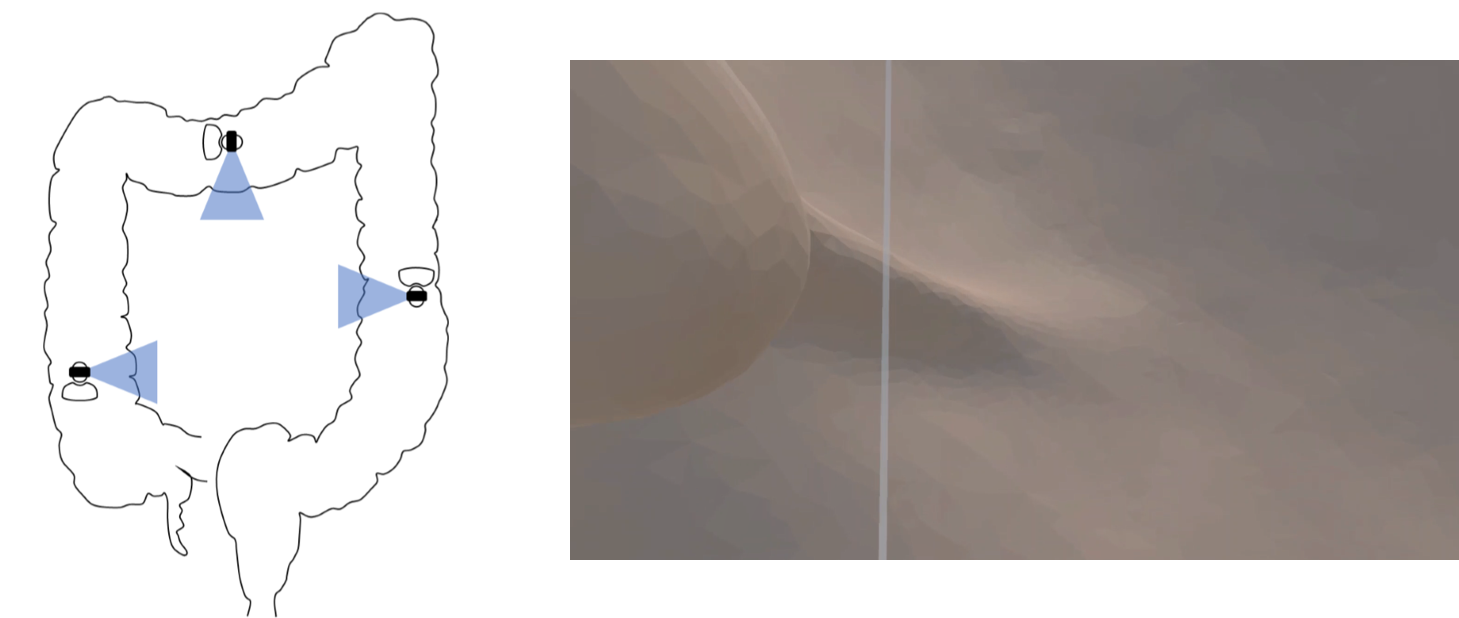}%
\label{fig:flyOver}
}
\subfigure[Elevator]{
\includegraphics[width=\columnwidth]{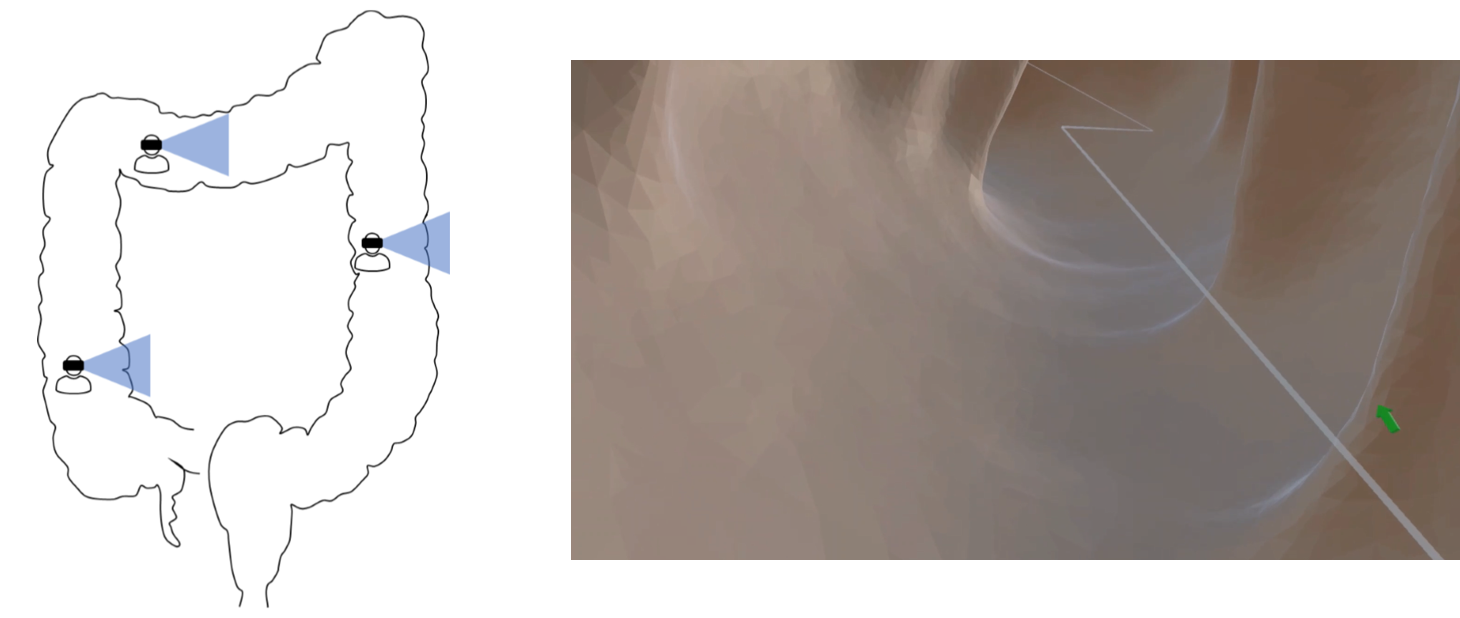}%
\label{fig:elevator}
}
\caption{Camera orientation schematics (left) and viewpoints (right) during (a) Fly-Through, (b) Fly-Over and (c) Elevator techniques. }
\label{fig:camOrientation}
\end{figure}

Each technique differs on how the user’s orientation is represented within the virtual environment. 
Identical to conventional CTC, the Fly-Through will make the user feel inside a cave. In this technique, the virtual camera follows the path without the need for users to move their head. They can, however, move their heads to see what is behind, below or above them. User orientation follows the centerline's direction, \changed{facing the center of the colon throughout the tortuous tubular structure} (\autoref{fig:camOrientation}(a)).
Differently from the traditional Fly-Over technique found in the literature~\cite{hassouna2006flyover} there is no need to split the colon in two halves and assign a virtual camera to each part. In this case, the inspection of the colon’s walls is done by users' head movement. The camera will automatically keep the perpendicular perspective in the eyes of the users, \changed{ facing the colon's wall,} while they can move their heads to analyze their surroundings \changed{as they move along the centerline} (\autoref{fig:camOrientation}(b)).
Finally, the Elevator technique does not change camera orientation, in order to match the user's real orientation (\autoref{fig:camOrientation}(c)). \changed{In this sense, the user will be facing the walls in ascending and descending segments of the colon, only facing the center of the colon whenever the centerline's orientation meets the user's real orientation.} Ultimately, this could reduce cybersickness during the VR trip, at the cost of increasing users' chances of losing the sense of direction.

\section{Evaluation}
\label{evaluation}

We compared three different camera travel techniques in order to investigate their potential effects on efficiency and diagnosis accuracy during CTC navigation: Fly-Over, Fly-Through and Elevator.
We used both quantitative and qualitative metrics to assess the ease of use, usefulness, efficiency and efficacy of each technique. 
Efficiency was measured based on task completion time, as efficacy corresponded to the success rate, i.e. the percentage of specific marks that were correctly identified. Through questionnaires (see \nameref{sec:appendix}) 
we assessed the subjective feeling of usefulness, ease of use and disorientation of all three techniques, as well as cybersickness~\cite{laviola2000discussion}. 

In addition, we conducted a semi-structured interview with two senior radiologists, in order to gain insights into the most relevant aspects in the diagnostic task, namely the impact of the accuracy rate and the task completion time in 
choosing a technique.

\subsection{Apparatus}

\begin{figure}[t]
\centering
\includegraphics[width=.9\columnwidth]{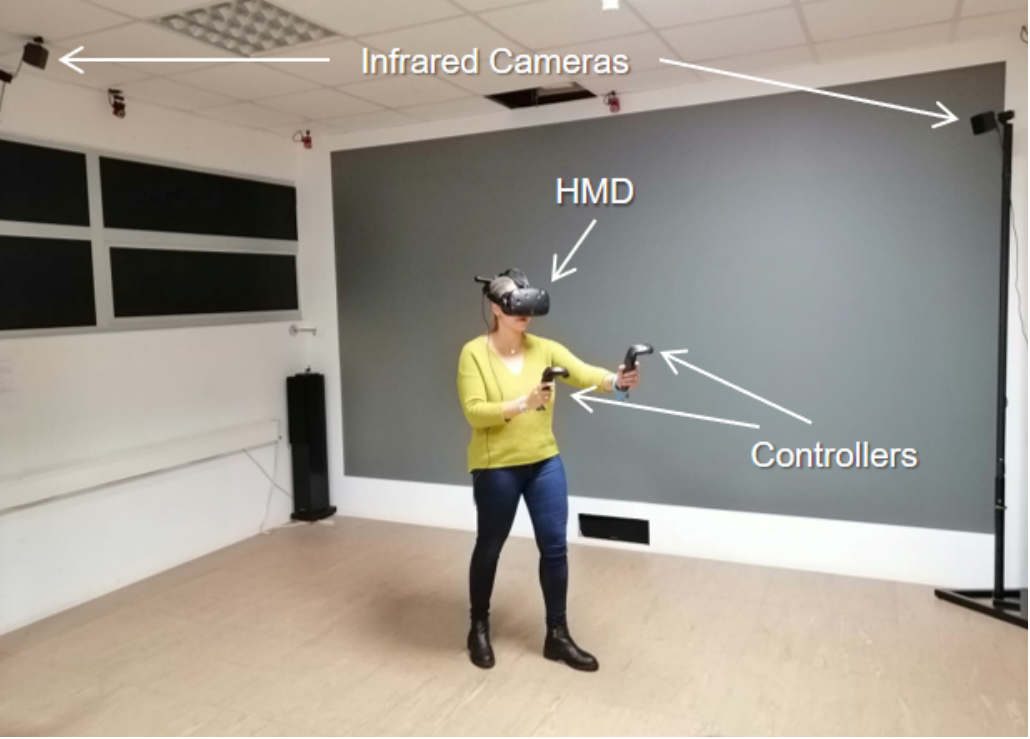}
\caption{VR setup of the immersive CTC interactive system.}
\label{fig:setup}
\end{figure}

Our setup relies on 
an off-the-shelf HTC Vive device. It consists of a binocular Head-Mounted Display, two game controllers and a Lighthouse Tracking System composed by two cameras with emitting pulsed infrared lasers that track all six degrees-of-freedom of head and handheld gear (\autoref{fig:setup}). 
The tracking system generates an acquisition volume that enables users to move freely within a 4.5x4.5x2.5 m\textsuperscript{3} space. 
We performed user tests 
using an Asus ROG G752VS Laptop with an Intel\textsuperscript\textregistered Core\texttrademark ~i7-6820HK Processor, 64GB RAM and NVIDIA GeForce GTX1070. The VR prototype runs at 60 frames per second.
All the code was developed in C\# using the SteamVR Plugin and Unity game engine (version 5.5.1f1).

\subsection{Participants}

Eighteen participants (13 male, 5 female) took part in our study, with ages between 18 and 25 years old (Mean = 21.94; Standard Deviation = 1.98). Most participants had an engineering background, namely Computer Science (38.89\%) and Biomedical Engineering (27.78\%). Most reported no previous experience in VR (66.6\%) or to use such systems less than once a month (27.78\%). One user reported to have claustrophobia.

\subsection{Methodology}

\begin{figure}[t]
\centering
\includegraphics[width=.6\columnwidth]{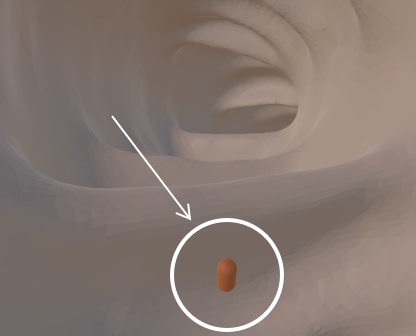}
\caption{Task performance: users were asked to find as many marks (orange) as they could while navigating inside the colon.}
\label{fig:prop}
\end{figure}

Participants were asked to complete a demographic questionnaire to survey their personal profile and previous experience regarding VR and medical tools.
This was followed by performing a training task with the technique they were assigned, to familiarize themselves both with the technique and the virtual environment.
After that, they performed the test task followed by a post-test questionnaire \changed{using a six-point Likert scale} to assess qualitative metrics, and a \changed{Simulator Sickness Questionnaire (SSQ) to assess cybersickness}. The task consisted in finding specific marks, in the form of orange \changed{3cm} capsules (\autoref{fig:prop}), which were placed in both easy and hard to find locations in the colon, to simulate the visibility of real lesions. Users were oblivious to the total amount of marks (20 marks per technique) spread throughout the colon. Instead, they were asked to find as many as they could, until they felt they had found them all.
This procedure was repeated for all three techniques, which were assigned according to a balanced latin-squares arrangement to avoid learning effects. 

\subsection{\changed{Radiologists' Qualitative Assessment}}

\changed{We conducted a semi-structured interview 
with two senior radiologists with 15 and 20 years of experience, both female and members of a multidisciplinary gastrointestinal oncology team. While both are familiar with CTC, only one performs CTC regularly, while the other only uses CTC for specific medical cases.
Due to pandemic restrictions, a single researcher conducted a remote session with both participants, in which both could freely share their opinions on the subject matter.
The session was video recorded and transcribed for data analysis.}

\section{Results}

In this section we present the results from our statistical analysis to evaluate quantitative and qualitative measures regarding the three techniques tested. \changed{To complement these results, we present the insights obtained from our interviews.}

For task completion time, success rate \changed{and SSQ scores}, a Shapiro-Wilk test revealed that the data were not normally distributed. We thus applied a Friedman non-parametric test for multiple comparisons and Wilcoxon Signed-Ranks post-hoc tests with a Bonferroni correction, setting a significance level at $p\leq0.017$. We also applied these tests  to Likert-scale data collected via questionnaires and cybersickness scores.

\begin{figure}[t]
    \centering
    \includegraphics[width=\columnwidth, trim=15 30 0 50, clip]{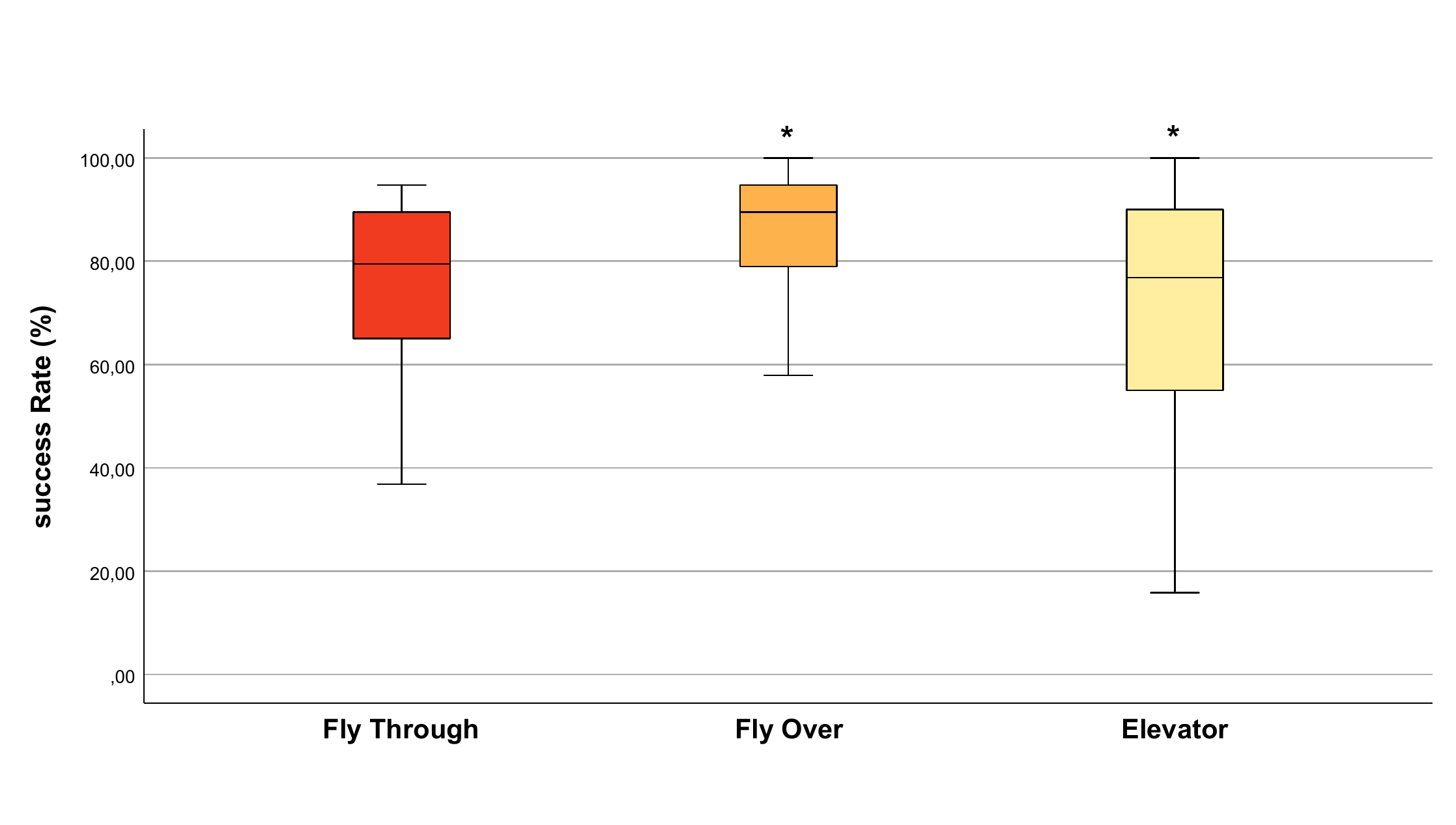}
    \caption{Success Rate for each condition: Fly-Through, Fly-Over and Elevator.* indicates statistical significance.}
    \label{fig:successRateBoxplot}
\end{figure}

There were 
significant differences in the success rate values depending on the 
technique used, $\chi^2(2)=7.600$, $p=0.022$.
Median (Interquartile Range) values for success rate using the Fly-Through, Fly-Over and Elevator techniques were 79.47 (26.25), 89.47 (18.42) and 76.84 (36.77), respectively (\autoref{fig:successRateBoxplot}). Post-hoc analysis 
showed a statistically significant increase of the success rate between the Elevator and the Fly-Over technique ($Z=-2.386, p=0.017$). However, there were no statistically significant differences between the Fly-Through and Fly-Over techniques ($Z = -2.345, p = 0.019$), nor between the Fly-Through and the Elevator ($Z = -0.734, p =0.463$).   

\begin{figure}[h!]
    \centering
    \includegraphics[width=\columnwidth,height=3.8cm, trim=15 30 0 20, clip]{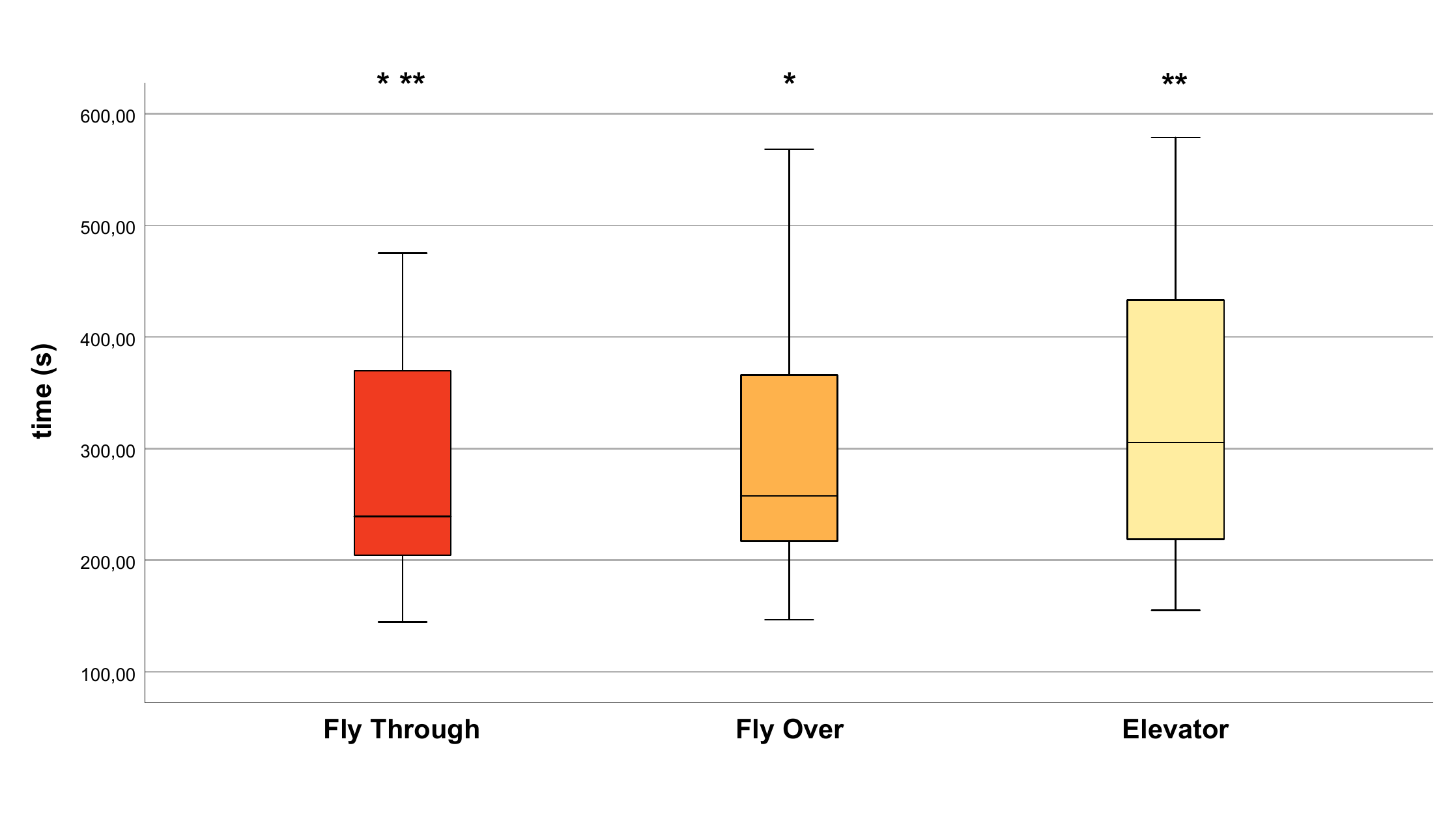}
    \caption{Task Completion Time by condition: Fly-Through, Fly-Over \& Elevator.* and ** indicate statistical significance.}
    \label{fig:timeBoxplot}
\end{figure}

\begin{figure}[t]
    \centering
    \includegraphics[width=\columnwidth, trim=81 525 75 80, clip]{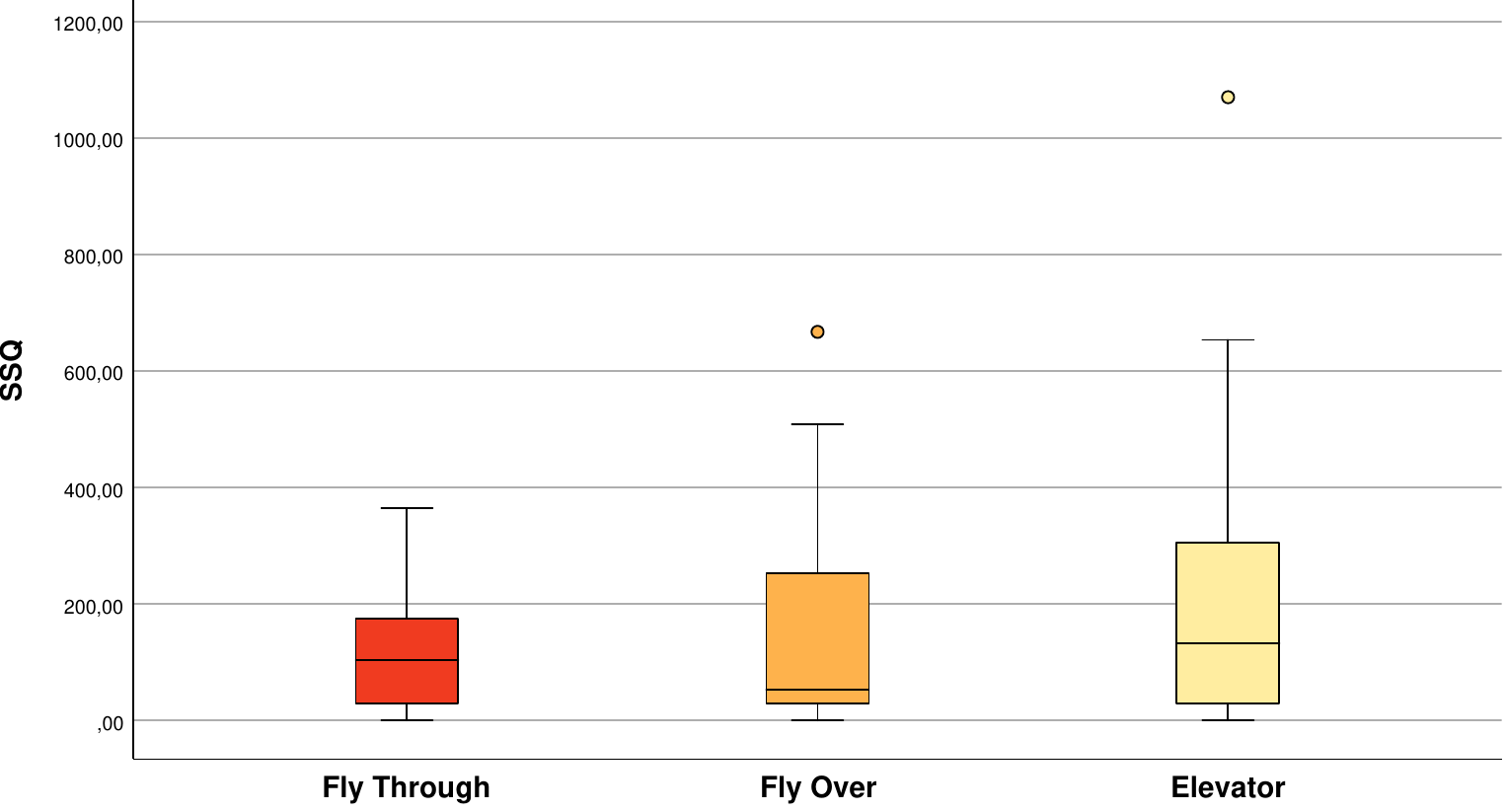}
    \caption{\changed{SSQ scores: Fly-Through, Fly-Over and Elevator.}}
    \label{fig:ssqBoxplot}
\end{figure}

Regarding task completion time, we found 
significant differences depending on the 
technique 
used, $\chi^2(2) = 10.333$, $p = 0.006$. Mean (standard deviation) task completion time values for the Fly-Through, Fly-Over and Elevator techniques were 273.91 (100.05), 305.53 (124.38) and 322.13 (135.49), respectively (\autoref{fig:timeBoxplot}).
Post-hoc analysis showed statistically significant decreases between Fly-Through and Fly-Over ($Z=-2.548, p=0.011$), as well as between the Fly-Through and the Elevator techniques ($Z = -2.548, p =0.011$). Still, there was no significant difference between the Fly-Over and the Elevator ($Z = -1.328, p = 0.184$). \changed{We also did not find significant differences between techniques regarding SSQ scores, $\chi^2(2)=4.875, p=0.087$ (\autoref{fig:ssqBoxplot}).}

As for qualitative metrics (\autoref{tab:questionnaires}), we found statistical significance in the perceived usefulness of the navigation technique (Q1) ($\chi^2(2)=7.35$ $p=0.025$).

Notably, users found Fly-Through more useful than the Fly-Over technique ($Z=-2.588$ $p=0.01$).
We also found statistical significance regarding the ease of understanding the direction of movement (Q2) ($\chi^2(2)=9.529$ $p=0.009$), but with no significance between pairs 
after performing 
a Bonferroni correction.
Finally, results indicate statistically significant differences in 
perceived disorientation ($\chi^2(2)=11.111$ $p=0.004$).
In 
effect, users felt less disoriented by the Fly-Through technique as compared either to  Elevator ($Z=-2.541$ $p=0.011$) or  Fly-Over ($Z=-2.634$ $p=0.008$) methods. 
\begin{table}[b]
\caption{Summary of the questionnaires split by question and technique (Fly-Through (\changed{FT}), Fly-Over (FO) and Elevator (EL)). Results are shown as Median (Interquartile Range).}
\resizebox{\columnwidth}{!}{
\begin{tabular}{@{}llll@{}}
\toprule
                                                                                                           & FT & FO & EL  \\ \midrule
Q1: Navigation was useful*                                                                                  & 6(1)       & 5(2)      & 6(2)      \\ \\
\begin{tabular}[c]{@{}l@{}}Q2: Direction of movement was easy to understand\end{tabular} & 6(0.25)    & 5(2)      & 5(2.25)   \\ \\
Q3: I was disoriented*                                                                                          & 1(1.5)     & 3.5(3.25) & 3(3.25)   \\ \\
Q4: It was easy to find the marks                                                                              & 5(2)       & 4.5(1.5)  & 5(2)      \\ \\
Q5: I felt that I found the same mark twice                                                                    & 2(2.25)    & 2.5(2.25) & 1.5(2.25) \\ \bottomrule
\end{tabular}
}
\label{tab:questionnaires}
\end{table}%

\subsection{\changed{Qualitative Assessment with Senior Radiologists}} 

\changed{From the radiologists point of view,\textit{ "CTC has some advantages and disadvantages over conventional optical colonoscopy. Our goal is to find polyps, just like gastroenterologists (who perform optical colonoscopies). We have to locate and measure polyps, so we are able to decide whether it is something too small, which needs to be watched over time, or if it is something that needs to be removed right away. That's the exam's main goal}".} 

\changed{When asked to characterize visualization challenges during endo-luminal navigation, radiologists agreed the colon is a complex structure with many anatomical barriers which directly impact the diagnostic task. In this sense, they mentioned that “\textit{The colon is a tortuous structure, which makes it difficult for the machines to fully capture the intestinal walls. This includes haustral folds, inflections and the colon as a tortuous structure by itself.}" The haustral folds in particular, require radiologists to navigate forward and backwards, in both prone and supine positions: \textit{“In this type of exam, we start by ascending from the rectum to the cecum, and then coming from the cecum back to the rectum to try to see both sides of the folds. This is part of the protocol, we do it both for the images obtained in prone (lying face down) and supine (lying face up) position. We try to see it in both directions to lessen the probability of missing something}”. Radiologists also pointed out "There is another type of pathology that results in a certain degree of inflammation of the colon, which is affected by diverticula and a lower bowel distension. This makes it harder to detect small polyps or any anomalies on the colon’s wall, and to differentiate the type of lesions found. Also, there are dirtier areas where bowel cleansing is not as effective, such as the right colon. Often, exams are not conclusive because people still have fecal matter inside their colon}”.

\changed{Finally, radiologists were asked about their perceptions on the importance of the accuracy rate and task completion time on diagnostic tasks. Firstly, they shared their considerations on task completion time: \textit{"Time is surely a relevant factor, it has become more and more important. After acquiring CTC data, we perform our diagnostic task at the workstation. We need to navigate the colon and if we find any lesions, we need to measure them and mark them. This never takes less than 15 minutes, even for an experienced radiologist"}. Still, they perceive accuracy rate as more relevant, not only for their performance, but also for patients' health. As they put it “\textit{I believe the accuracy rate is more important, isn’t it? As long as the technique is accurate and enables us to perform the medical diagnosis, it can be more time consuming. If it was faster, it would be better, but we will use it anyway. We will not discard it for that}”. Also, professionals highlighted the importance of the accuracy rate by saying that “\textit{In any exam, the main goal is to achieve a full pathology detection and achieve the highest accuracy rate, close to 100\%. We also know that it is hard, not just with CTC, but also conventional optical colonoscopy. I can’t recall the current accuracy rate of CTC performed in a clean colon, in an exam that is considered well done, but it is very close to the other (optical colonoscopy). Sure that if we miss any lesions, we also miss the opportunity of removing them and it may be the case they eventually transform into a cancerous lesion. On the other hand, if we believe that we detected a cancerous lesion, which turns out to be a false lesion, and not a polypoid lesion, then we will be sending the patient to undergo optical colonoscopy to potentially locate and remove the polypoid lesion in a more invasive procedure. That can happen in any exam!}}”.

\section{Discussion}

\changed{We have presented a study on three different techniques for locomotion in VR Colonoscopy using head-mounted displays. While fly-through and fly-over have been previously studied in desktop applications, we introduced a third novel technique (elevator) and compared the three via empirical tests with novice users.} Overall, Fly-Through has proven to be the best technique for immersive colonoscopy navigation in the user tests we conducted. Indeed, we found it to be the most efficient option, according to task completion times, besides being considered the most useful (Q1), easy to use (Q2) and less disorienting (Q3) by the subjects. Even though Fly-Over seemingly produced higher success rates, there were no statistical differences that could support its use over the Fly-Through, since the significant increase in task completion times would likely offset those gains. \changed{However, from a medical perspective, significant improvements in accuracy would outweigh longer task completion times, which would justify investigating this technique further.}  

Additionally, users reported higher disorientation while using the Fly-Over technique. Such results may be attributed to the fact that Fly-Over had subjects facing the colon walls most of the time. 
Orienting the camera at a direction perpendicular to displacement 
severely hampered their general perception of the tubular structure of the colon and the path they were following. This, combined with marks located behind their 
backs, which forced subjects to inspect the structure in several directions to try and find them, ultimately caused their disorientation. Such results suggest that the Fly-Over technique may be improved by 
devising new means and interaction techniques for clinicians to visualize structures on their back without the need to physically turn.
By doing this they could combine both the observed effectiveness of the Fly-Over technique with more efficient means to 
support camera travel in immersive CTC navigation.
Finally, the Elevator technique was the least favored option for navigating the virtual environment. 
This may be due to the search strategy adopted by most, which had to change after each abrupt movement caused by the natural inflections of the colon's structure. That may also explain why this technique turned out to be the least efficient when compared to the other two approaches, as users required more time to adapt and adjust their orientation whenever the camera direction changed. \changed{Surprisingly, we could not find significant differences in terms of cybersickness reported by subjects.}

\section{Conclusions}

In this paper, we 
study Fly-Through and Fly-Over techniques in immersive VR CTC, 
in terms of efficiency, ease of use, usefulness and effectiveness. We also compared these to the Elevator, a novel technique in this domain that combines both approaches to make virtual orientation match the user's direction of movement throughout navigation. Our results show that Fly-Through is still the most efficient and easy to use technique for immersive VR CTC. We found the Elevator technique to be less effective and efficient than both Fly-Through and Fly-Over methods, but less disorienting than the Fly-Over approach.
This can be explained by the need to physically turn one's body to effectively scan the colon structure in all directions. Still, this limitation did not affect task effectiveness, as in the Fly-Over technique users 
could achieve higher success rates when finding specific marks along the colonic structure.
Thus, the Fly-Over would be the technique of choice in order to provide a more accurate analysis, and produce enhanced readings, as it helps people to identify lesions even in difficult-to-scan locations despite being a more time-consuming procedure. Indeed our experience suggests that each interaction technique could be useful in its own right, Fly-Through being most adequate to scan the colon in a quick preview, while Fly-Over would likely enable more reliable and comprehensive readings by clinicians, \changed{which would potentially make it their technique of choice}. 

Still, our study had two main limitations. First, our experimental task only aimed
at reflecting the real clinical task to a certain extent, i.e., limited lesion visibility caused by the anatomical properties of the colon, while orange capsules may significantly differ from lesions such as polyps. Second, our participants had no clinical background, which may impact the selection of the ideal navigation technique to perform immersive VR CTC analysis. Future work will include validating such conclusions with medical professionals and
using more generic flying techniques and possibly additional interface modalities  to improve diagnostic and generalize our results to more cave- and tunnel-like structures.


\begin{acknowledgements}

This work was supported by Fundação para a Ciência e a Tecnologia, Portugal, through grant numbers UIDB/50021/2020 and SFRH/BD/136212/2018, and by MBIE grant ILF-VUW1901, New Zealand. \changed {We are grateful to Dr. Isabel Nobre and Dr. Sandra Sousa from the Imagiology Service of \textit{Hospital Lusíadas Lisboa} for their professional assessment and comments}. The authors would also like to thank Pedro Borges for his contributions during his MSc thesis work.
\end{acknowledgements}

%
\section*{Conflict of interest}

The authors declare that they have no conflict of interest.

\bibliographystyle{spmpsci}      
\balance
\bibliography{main}   

\begin{thebibliography}{10}
\providecommand{\url}[1]{{#1}}
\providecommand{\urlprefix}{URL }
\expandafter\ifx\csname urlstyle\endcsname\relax
  \providecommand{\doi}[1]{DOI~\discretionary{}{}{}#1}\else
  \providecommand{\doi}{DOI~\discretionary{}{}{}\begingroup
  \urlstyle{rm}\Url}\fi

\bibitem{Ferlay2015}
Ferlay, J., Soerjomataram, I., Dikshit, R., Eser, S., Mathers, C., Rebelo, M.,
  Parkin, D.M., Forman, D., Bray, F.: {Cancer incidence and mortality
  worldwide: Sources, methods and major patterns in GLOBOCAN 2012}.
\newblock International Journal of Cancer \textbf{136}(5), E359--E386 (2015).
\newblock \doi{10.1002/ijc.29210}

\bibitem{yao2010reversible}
Yao, J., Chowdhury, A.S., Aman, J., Summers, R.M.: Reversible projection
  technique for colon unfolding.
\newblock IEEE Transactions on Biomedical engineering \textbf{57}(12),
  2861--2869 (2010)

\bibitem{mirhosseini2014}
Mirhosseini, K., Sun, Q., Gurijala, K.C., Laha, B., Kaufman, A.E.: {Benefits of
  3D immersion for virtual colonoscopy}.
\newblock In: 2014 IEEE VIS International Workshop on 3DVis (3DVis), pp. 75--79
  (2014).
\newblock \doi{10.1109/3DVis.2014.7160105}

\bibitem{Schuchardt07}
Schuchardt, P., Bowman, D.A.: The benefits of immersion for spatial
  understanding of complex underground cave systems.
\newblock In: Proceedings of the 2007 ACM Symposium on Virtual Reality Software
  and Technology, VRST ’07, p. 121–124. Association for Computing
  Machinery, New York, NY, USA (2007).
\newblock \doi{10.1145/1315184.1315205}

\bibitem{bowman1997travel}
Bowman, D.A., Koller, D., Hodges, L.F.: Travel in immersive virtual
  environments: An evaluation of viewpoint motion control techniques.
\newblock In: Proceedings of IEEE 1997 Annual International Symposium on
  Virtual Reality, pp. 45--52. IEEE (1997)

\bibitem{elmqvist2006navigation}
Elmqvist, N., Tsigas, P.: On navigation guidance for exploration of 3d
  environments.
\newblock G{\"o}teborg, Sweden  (2006)

\bibitem{elmqvist2007tour}
Elmqvist, N., Tudoreanu, M.E., Tsigas, P.: Tour generation for exploration of
  3d virtual environments.
\newblock In: Proceedings of the 2007 ACM Symposium on Virtual Reality Software
  and Technology, VRST ’07, p. 207–210. Association for Computing
  Machinery, New York, NY, USA (2007).
\newblock \doi{10.1145/1315184.1315224}

\bibitem{bartz2005virtual}
Bartz, D.: Virtual endoscopy in research and clinical practice.
\newblock Computer Graphics Forum \textbf{24}(1), 111--126 (2005).
\newblock \doi{10.1111/j.1467-8659.2005.00831.x}

\bibitem{he2001reliable}
He, T., Hong, L., Chen, D., Liang, Z.: Reliable path for virtual endoscopy:
  ensuring complete examination of human organs.
\newblock IEEE Transactions on Visualization and Computer Graphics
  \textbf{7}(4), 333--342 (2001).
\newblock \doi{10.1109/2945.965347}

\bibitem{huang2005teniae}
Huang, A., Roy, D., Franaszek, M., Summers, R.M.: Teniae coli guided navigation
  and registration for virtual colonoscopy.
\newblock In: IEEE Conference on Visualization, VIS 05, pp. 279--285. IEEE
  (2005).
\newblock \doi{10.1109/VISUAL.2005.1532806}

\bibitem{hong19953d}
Hong, L., Kaufman, A., Wei, Y.C., Viswambharan, A., Wax, M., Liang, Z.: 3d
  virtual colonoscopy.
\newblock In: Biomedical Visualization, 1995. Proceedings., pp. 26--32. IEEE
  (1995)

\bibitem{hassouna2006flyover}
Hassouna, M.S., Farag, A.A., Falk, R.: Virtual fly-over: A new visualization
  technique for virtual colonoscopy.
\newblock In: MICCAI 2006, pp. 381--388. Springer Berlin Heidelberg, Berlin,
  Heidelberg (2006)

\bibitem{Robinett1992}
Robinett, W., Holloway, R.: Implementation of flying, scaling and grabbing in
  virtual worlds.
\newblock In: Symposium On Interactive 3d Graphics, pp. 189--192. ACM Press
  (1992).
\newblock \doi{http://doi.acm.org/10.1145/147156.147201}

\bibitem{lopes2018interaction}
Lopes, D.S., Medeiros, D., Paulo, S.F., Borges, P.B., Nunes, V., Mascarenhas,
  V., Veiga, M., Jorge, J.A.: Interaction techniques for immersive ct
  colonography: A professional assessment.
\newblock In: MICCAI 2018, pp. 629--637. Springer (2018).
\newblock \doi{10.1007/978-3-030-00934-2_70}

\bibitem{Haker2000NondistortingFM}
Haker, S., Angenent, S.B., Tannenbaum, A.R., Kikinis, R.: Nondistorting
  flattening maps and the 3-d visualization of colon ct images.
\newblock IEEE Transactions on Medical Imaging \textbf{19}, 665--670 (2000)

\bibitem{wang2015novel}
Wang, H., Chen, Y., Li, L., Pan, H., Gu, X., Liang, Z.: A novel colon wall
  flattening model for computed tomographic colonography: method and
  validation.
\newblock Computer Methods in Biomechanics and Biomedical Engineering: Imaging
  \& Visualization \textbf{3}(4), 213--221 (2015)

\bibitem{vos2003three}
Vos, F.M., van Gelder, R.E., Serlie, I.W., Florie, J., Nio, C.Y., Glas, A.S.,
  Post, F.H., Truyen, R., Gerritsen, F.A., Stoker, J.: Three-dimensional
  display modes for ct colonography: conventional 3d virtual colonoscopy versus
  unfolded cube projection.
\newblock Radiology \textbf{228}(3), 878--885 (2003)

\bibitem{codd2011virtual}
Codd, A.M., Choudhury, B.: Virtual reality anatomy: Is it comparable with
  traditional methods in the teaching of human forearm musculoskeletal anatomy?
\newblock Anatomical sciences education \textbf{4}(3), 119--125 (2011)

\bibitem{de2011progress}
De~Visser, H., Watson, M.O., Salvado, O., Passenger, J.D.: Progress in virtual
  reality simulators for surgical training and certification.
\newblock Medical Journal of Australia \textbf{194}, S38--S40 (2011)

\bibitem{vosburgh2013surgery}
Vosburgh, K.G., Golby, A., Pieper, S.D.: Surgery, virtual reality, and the
  future.
\newblock Studies in health technology and informatics \textbf{184}, vii (2013)

\bibitem{shanmugan2014virtual}
Shanmugan, S., Leblanc, F., Senagore, A.J., Ellis, C.N., Stein, S.L., Khan, S.,
  Delaney, C.P., Champagne, B.J.: Virtual reality simulator training for
  laparoscopic colectomy: what metrics have construct validity?
\newblock Diseases of the Colon \& Rectum \textbf{57}(2), 210--214 (2014)

\bibitem{king2016immersive}
King, F., Jayender, J., Bhagavatula, S.K., Shyn, P.B., Pieper, S., Kapur, T.,
  Lasso, A., Fichtinger, G.: An immersive virtual reality environment for
  diagnostic imaging.
\newblock Journal of Medical Robotics Research \textbf{1}(01), 1640003 (2016).
\newblock \doi{10.1142/S2424905X16400031}

\bibitem{sousa2017vrrrroom}
Sousa, M., Mendes, D., Paulo, S., Matela, N., Jorge, J., Lopes, D.S.:
  {VRRRRoom: Virtual Reality for Radiologists in the Reading Room}.
\newblock In: Proceedings of the 2017 CHI Conference on Human Factors in
  Computing Systems, pp. 4057--4062. ACM (2017).
\newblock \doi{10.1145/3025453.3025566}

\bibitem{wirth2018evaluation}
Wirth, M., Gradl, S., Sembdner, J., Kuhrt, S., Eskofier, B.M.: Evaluation of
  interaction techniques for a virtual reality reading room in diagnostic
  radiology.
\newblock In: The 31st Annual ACM Symposium on User Interface Software and
  Technology, pp. 867--876. ACM (2018)

\bibitem{medeiros2019magic}
Medeiros, D., Sousa, M., Raposo, A., Jorge, J.: Magic carpet: Interaction
  fidelity for flying in vr.
\newblock IEEE transactions on visualization and computer graphics
  \textbf{26}(9), 2793--2804 (2019)

\bibitem{chaudhuri2004efficient}
Chaudhuri, P., Khandekar, R., Sethi, D., Kalra, P.: An efficient central path
  algorithm for virtual navigation.
\newblock In: Proceedings Computer Graphics International, 2004., pp. 188--195.
  IEEE (2004)

\bibitem{cheng2014augmented}
Cheng, I., Shen, R., Moreau, R., Brizzi, V., Rossol, N., Basu, A.: An augmented
  reality framework for optimization of computer assisted navigation in
  endovascular surgery.
\newblock In: 2014 36th Annual International Conference of the IEEE Engineering
  in Medicine and Biology Society, pp. 5647--5650. IEEE (2014)

\bibitem{noser2003automatic}
Noser, H., Stern, C., Stucki, P.: Automatic path searching for interactive
  navigation support within virtual medical 3d objects.
\newblock In: International Congress Series, vol. 1256, pp. 29--34. Elsevier
  (2003)

\bibitem{aguilar2017rrt}
Aguilar, W.G., Abad, V., Ruiz, H., Aguilar, J., Aguilar-Castillo, F.: Rrt-based
  path planning for virtual bronchoscopy simulator.
\newblock In: International Conference on Augmented Reality, Virtual Reality
  and Computer Graphics, pp. 155--165. Springer (2017)

\bibitem{haigron2004depth}
Haigron, P., Bellemare, M.E., Acosta, O., Goksu, C., Kulik, C., Rioual, K.,
  Lucas, A.: Depth-map-based scene analysis for active navigation in virtual
  angioscopy.
\newblock IEEE Transactions on Medical Imaging \textbf{23}(11), 1380--1390
  (2004)

\bibitem{huang2006synchronous}
Huang, A., Summers, R.M., Roy, D.: Synchronous navigation for ct colonography.
\newblock In: Medical Imaging 2006: Physiology, Function, and Structure from
  Medical Images, vol. 6143, p. 614315. International Society for Optics and
  Photonics (2006)

\bibitem{pareek2018survey}
Pareek, T.G., Mehta, U., Gupta, A., et~al.: A survey: Virtual reality model for
  medical diagnosis.
\newblock Biomedical and Pharmacology Journal \textbf{11}(4), 2091--2100 (2018)

\bibitem{venson2016medical}
Venson, J.E., Berni, J., Maia, C.S., da~Silva, A.M., d'Ornelas, M., Maciel, A.:
  Medical imaging vr: Can immersive 3d aid in diagnosis?
\newblock In: Proceedings of the 22nd ACM Conference on Virtual Reality
  Software and Technology, pp. 349--350 (2016)

\bibitem{mirhosseini2019immersive}
Mirhosseini, S., Gutenko, I., Ojal, S., Marino, J., Kaufman, A.: Immersive
  virtual colonoscopy.
\newblock IEEE transactions on visualization and computer graphics
  \textbf{25}(5), 2011--2021 (2019)

\bibitem{randall2015oculus}
Randall, D., Metherall, P., Bardhan, K.D., Spencer, P., Gillott, R.,
  de~Noronha, R., Fenner, J.W.: The oculus rift virtual colonoscopy:
  introducing a new technology and initial impressions.
\newblock Journal of Biomedical Graphics and Computing \textbf{6}(1) (2015).
\newblock \doi{10.5430/jbgc.v6n1p34}

\bibitem{tciaCTC}
Smith, K., Clark, K., Bennett, W., Nolan, T., Kirby, J., Wolfsberger, M.,
  Moulton, J., Vendt, B., Freymann, J.: Data from {CT{\_}COLONOGRAPHY}.
\newblock Retrieved April 23, 2017
  \url{http://doi.org/10.7937/K9/TCIA.2015.NWTESAY1} (2015)

\bibitem{ribeiro20093}
Ribeiro, N., Fernandes, P., Lopes, D., Folgado, J., Fernandes, P.: 3-d solid
  and finite element modeling of biomechanical structures—a software
  pipeline.
\newblock In: 7th EUROMECH Solid Mechanics Conference (2009)

\bibitem{tagliasacchi2012mean}
Tagliasacchi, A., Alhashim, I., Olson, M., Zhang, H.: Mean curvature skeletons.
\newblock Computer Graphics Forum \textbf{31}(5), 1735--1744 (2012).
\newblock \doi{10.1111/j.1467-8659.2012.03178.x}

\bibitem{laviola2000discussion}
LaViola~Jr, J.J.: A discussion of cybersickness in virtual environments.
\newblock ACM SIGCHI Bulletin \textbf{32}(1), 47--56 (2000)

\end{thebibliography}

\clearpage\onecolumn\section*{\changed{Appendix}}
\label{sec:appendix}

\subsection*{\changed{Navigation Experience Questionnaire}}
Navigation technique: Fly-Through $\square \quad$ Fly-Over $\square \quad$ Elevator $\square$ \hfill \\
\hfill \\
\noindent
Q1. Was this type of navigation useful? \hfill \\
$Totally \> Disagree \quad 1\square \quad 2\square \quad 3\square \quad 4\square \quad 5\square \quad 6\square \quad Totally\> Agree$ \hfill \\
\hfill \\
Q2. Was it easy to understand the direction of movement?\hfill \\
$Totally \> Disagree \quad 1\square \quad 2\square \quad 3\square \quad 4\square \quad 5\square \quad 6\square \quad Totally\> Agree$ \hfill \\
\hfill \\
Q3. Did you feel disoriented?\hfill \\
$Totally \> Disagree \quad 1\square \quad 2\square \quad 3\square \quad 4\square \quad 5\square \quad 6\square \quad Totally\> Agree$ \hfill \\
\hfill \\
Q4. Was it easy to find the capsules?\hfill \\
$Totally \> Disagree \quad 1\square \quad 2\square \quad 3\square \quad 4\square \quad 5\square \quad 6\square \quad Totally\> Agree$ \hfill \\
\hfill \\
Q5. Do you think you marked the same capsule twice?\hfill \\
$Totally \> Disagree \quad 1\square \quad 2\square \quad 3\square \quad 4\square \quad 5\square \quad 6\square \quad Totally\> Agree$

\subsection*{\changed{User Experience Questionnaire}}
Q1. Did you find the way of moving (forward or backwards) suitable for the task? \hfill \\
$Totally \> Disagree \quad 1\square \quad 2\square \quad 3\square \quad 4\square \quad 5\square \quad 6\square \quad Totally\> Agree$ \hfill \\
\hfill \\
Q2. Was it ease to move?\hfill \\
$Totally \> Disagree \quad 1\square \quad 2\square \quad 3\square \quad 4\square \quad 5\square \quad 6\square \quad Totally\> Agree$ \hfill \\
\hfill \\
Q3. Is it easy to remember how to move?\hfill \\
$Totally \> Disagree \quad 1\square \quad 2\square \quad 3\square \quad 4\square \quad 5\square \quad 6\square \quad Totally\> Agree$ \hfill \\
\hfill \\
Q4. Did you find the way of marking capsules was suitable for the task?\hfill \\
$Totally \> Disagree \quad 1\square \quad 2\square \quad 3\square \quad 4\square \quad 5\square \quad 6\square \quad Totally\> Agree$ \hfill \\
\hfill \\
Q5. Was it ease to perform the action of marking a capsule?\hfill \\
$Totally \> Disagree \quad 1\square \quad 2\square \quad 3\square \quad 4\square \quad 5\square \quad 6\square \quad Totally\> Agree$ \hfill \\
\hfill \\
Q6. Is it easy to remember how to mark a capsule?\hfill \\
$Totally \> Disagree \quad 1\square \quad 2\square \quad 3\square \quad 4\square \quad 5\square \quad 6\square \quad Totally\> Agree$ \hfill \\

\noindent
Do you have any comments and/or suggestions about this approach?

\subsection*{\changed{Simulator Sickness Questionnaire}}

Before navigation $\square \quad$ After navigation $\square \quad$ \hfill \\
\hfill \\
Navigation technique: Fly-Through $\square \quad$ Fly-Over $\square \quad$ Elevator $\square$ \hfill \\

\noindent
Indicate how much each symptom below is affecting you now.\hfill \\

\noindent
General discomfort \hfill \\
$Not \> at \> all \quad \square \quad Mild\square \quad Moderate\square \quad Severe\square \quad$ \hfill \\
\hfill \\
Fatigue \hfill \\
$Not \> at \> all \quad \square \quad Mild\square \quad Moderate\square \quad Severe\square \quad$ \hfill \\
\hfill \\
Headache \hfill \\
$Not \> at \> all \quad \square \quad Mild\square \quad Moderate\square \quad Severe\square \quad$ \hfill \\
\hfill \\
Eyestrain \hfill \\
$Not \> at \> all \quad \square \quad Mild\square \quad Moderate\square \quad Severe\square \quad$ \hfill \\
\hfill \\
Difficulty focusing \hfill \\
$Not \> at \> all \quad \square \quad Mild\square \quad Moderate\square \quad Severe\square \quad$ \hfill \\
\hfill \\
Increased salivation \hfill \\
$Not \> at \> all \quad \square \quad Mild\square \quad Moderate\square \quad Severe\square \quad$ \hfill \\
\hfill \\
Sweating \hfill \\
$Not \> at \> all \quad \square \quad Mild\square \quad Moderate\square \quad Severe\square \quad$ \hfill \\
\hfill \\
Nausea \hfill \\
$Not \> at \> all \quad \square \quad Mild\square \quad Moderate\square \quad Severe\square \quad$ \hfill \\
\hfill \\
Difficulty concentrating \hfill \\
$Not \> at \> all \quad \square \quad Mild\square \quad Moderate\square \quad Severe\square \quad$ \hfill \\
\hfill \\
Fullness of head \hfill \\
$Not \> at \> all \quad \square \quad Mild\square \quad Moderate\square \quad Severe\square \quad$ \hfill \\
\hfill \\
Blurred vision \hfill \\
$Not \> at \> all \quad \square \quad Mild\square \quad Moderate\square \quad Severe\square \quad$ \hfill \\
\hfill \\
Dizziness (eyes open) \hfill \\
$Not \> at \> all \quad \square \quad Mild\square \quad Moderate\square \quad Severe\square \quad$ \hfill \\
\hfill \\
Dizziness (eyes closed) \hfill \\
$Not \> at \> all \quad \square \quad Mild\square \quad Moderate\square \quad Severe\square \quad$ \hfill \\
\hfill \\
Vertigo \hfill \\
$Not \> at \> all \quad \square \quad Mild\square \quad Moderate\square \quad Severe\square \quad$ \hfill \\
\hfill \\
Stomach awareness \hfill \\
$Not \> at \> all \quad \square \quad Mild\square \quad Moderate\square \quad Severe\square \quad$ \hfill \\
\hfill \\
Burping \hfill \\
$Not \> at \> all \quad \square \quad Mild\square \quad Moderate\square \quad Severe\square \quad$ \hfill \\
\hfill \\
Other:

\end{document}